\documentclass[preprint,12pt,3p]{elsarticle}
\usepackage{epsfig,amssymb,amsfonts,bm}
\biboptions{sort&compress}

\newcommand{\be}{\begin{equation}}
\newcommand{\ee}{\end{equation}}
\newcommand{\bea}{\begin{eqnarray}}
\newcommand{\eea}{\end{eqnarray}}
\newcommand{\vep}{{\bm p}}
\newcommand{\veq}{{\bm q}}
\newcommand{\vek}{{\bm k}}
\newcommand{\ds}{\displaystyle}
\newcommand{\0}{^{\rm{ph}}}
\newcommand{\oo}[1]{(#1^{\rm ph})^2}
\DeclareMathSymbol{\varGamma}{\mathord}{letters}{"00}

\journal{Physics Letters B}

\begin{document}

\begin{frontmatter}

\title{Quark mass dependence of the $X(3872)$ binding energy}

\author[1,2]{V. Baru}
\author[1]{E. Epelbaum}
\author[1,2]{A. A. Filin}
\author[3]{C. Hanhart}
\author[3,4]{U.-G. Mei\ss ner}
\author[2,5,6]{A. V. Nefediev}

\address[1]{Institut f\"ur Theoretische Physik II, Ruhr-Universit\"at Bochum, D-44780 Bochum, Germany}
\address[2]{Institute for Theoretical and Experimental Physics, B. Cheremushkinskaya 25, 117218 Moscow, Russia}
\address[3]{Forschungszentrum J\"ulich, Institute for Advanced Simulation, Institut f\"ur Kernphysik and
J\"ulich Center for Hadron Physics, D-52425 J\"ulich, Germany}
\address[4]{Helmholtz-Institut f\"ur Strahlen- und Kernphysik and 
Bethe Center for Theoretical Physics, Universit\"{a}t Bonn, D-53115 Bonn, Germany} 
\address[5]{National Research Nuclear University MEPhI, 115409, Moscow, Russia}
\address[6]{Moscow Institute of Physics and Technology, 141700, Dolgoprudny, Moscow Region, Russia}

\begin{abstract}
We explore the quark-mass dependence of the pole position of the $X(3872)$ state
within the molecular picture. The calculations are performed within the framework of a nonrelativistic
Faddeev-type three-body equation for the $D\bar{D}\pi$ system in the $J^{PC}=1^{++}$
channel. The $\pi D$ interaction is parametrised via a $D^*$ pole, and a three-body force is
included to render the equations well defined. Its strength is adjusted such that the
$X(3872)$ appears as a $D\bar{D}^*$ bound state 0.5~MeV below the neutral threshold.
We find that the trajectory of the $X(3872)$ depends strongly on the assumed quark-mass
dependence of the short-range interactions which can be determined in
future lattice QCD calculations. At the same time we are able
to provide nontrivial information on the chiral extrapolation in the $X$
channel.
\end{abstract}


\end{frontmatter}

\section{Introduction}

A decade ago, a new era in hadronic spectroscopy began with the observation by
the Belle Collaboration
of the charmonium-like meson $X(3872)$ \cite{Xobservation} which was quickly confirmed
by many other collaborations, see Ref.~\cite{QWGreview} for a recent
review article. In fact, such a state was predicted to exist as a $D\bar{D}^*$ bound
system, analogous to the deuteron, long before its discovery, based on
scattering calculations using a one-pion-exchange (OPE)
potential~\cite{molecule1,molecule2}.
This state possesses a number of intriguing features which
make it an attractive object for both experimental and theoretical studies. In particular, it is seen  with
approximately equal branching fractions  in the modes $\pi^+\pi^- J/\psi$ \cite{Belle2pi} and $\pi^+\pi^-\pi^0 J/\psi$
\cite{Babar3pi},   which  indicates a strong isospin violation in the $X$ decay. 
Recently, the quantum numbers of the $X(3872)$ were measured to be $1^{++}$~\cite{LHCb}, consistent
with its being an $S$-wave $D\bar D^*$\footnote{This shorthand notation is used for
the proper $C$-parity eigenstate.} bound
system~\cite{Braaten:2007dw,Canham:2009zq,Kalashnikova:2009gt,Kalashnikova:2010zz}.
This interpretation finds support especially in the very small binding energy, $E_B<1$ MeV, with respect to the 
$D^0\bar{D}^{*0}$ channel~\cite{PDG}.

While some authors claim that the iterated OPE potential alone is strong enough to form the
$X(3872)$~\cite{molecule3,molecule4,ThCl,Wang:2013kva}, others come to an opposite conclusion,
see, for example, \cite{Liu:2008fh,Kalashnikova:2012qf}. Different (complementary or alternative to the OPE)
short-range $D\bar{D}^*$ mechanisms  in the $X$ are considered, for instance, in
\cite{Liu:2008tn,Ding:2009vj,coupled2,coupled3}. It has to be noticed, however, that  pions are treated as
static in most models for the $X$ involving OPE. Furthermore, as stressed in \cite{suzuki}, since the 
$X$ resides only $7$~MeV above the $D\bar{D}\pi$
threshold, the presence of the corresponding cut might weaken the pion potential considerably. 

In order to better understand the role of the OPE interaction  including that of the three-body cut  as well as
to use a treatment with minimal bias, the $X(3872)$ was studied in \cite{Baru:2011rs} 
within a three-body scattering Faddeev-type formalism for the $D\bar{D}\pi$ system. The $\pi D$
interactions were parametrised via the $D^*$ pole.
In addition to the OPE, a $D\bar{D}^*$ contact term with a strength $C_0$ was also included to parametrise  
unknown short-distance physics. The resulting interaction was   iterated to all
orders through the solution of the integral equation. The equations were solved in momentum space using a 
sharp cut-off prescription. The scale-dependence of the contact term was determined by requiring
the equations to have a pole at a fixed binding energy of $0.5$ MeV for a wide range of different cut-offs $\Lambda$. 
It was found, in particular, that $C_0$ varies strongly with $\Lambda$ taking values between $\pm \infty$.
Interestingly,
for $\Lambda\simeq 1$ GeV, the contact term turns out to be very small which might explain the results of 
\cite{molecule3,molecule4,ThCl,Wang:2013kva}. However, it follows from \cite{Baru:2011rs} that these
findings are  scheme-dependent.
The low-energy dynamics of the $D\bar D^*$ system was also investigated within a
pionless EFT \cite{AlFiky:2005jd} as well as the so-called X-EFT approach (see, for example,
\cite{Fleming:2007rp,Voloshin:2003nt}) that assumes that pions can be treated perturbatively. 

In this paper we apply the formalism developed in
\cite{Baru:2011rs} to explore the behaviour of the binding energy $E_B$ as
the quark masses go away from their physical value. Since the pion mass
enters our calculations explicitly and the pion mass squared is directly proportional
to the quark mass then, equivalently, in the following we talk about the pion mass dependence of $E_B$. To this end, we
treat all masses and coupling constants as functions of the pion mass
$m_\pi$ and define their physical values (labelled as ``ph'') as those
which correspond to the physical pion mass $m_\pi\0$ or, equivalently,
the physical values of the light quark masses. We refer to this case as to the physical limit or physical point. We then
perform an
expansion of all such quantities in terms of the parameter $\delta 
m_\pi/{\rm M}$, where the small scale is $\delta m_\pi=m_\pi-m_\pi\0$, while the
large scale ${\rm M}$ is given by a typical hadronic scale such
as, for instance, the $D$- or $D^*$-meson mass or the chiral symmetry breaking
scale $\sim 4\pi f_\pi$,  with $f_\pi$ the pion  decay  constant. In either case, ${\rm M}$ appears to be 
of order $1$~GeV so that  keeping both  the leading term in the expansion  $\propto \delta m_\pi^2/{\rm M}^2$ and  the
leading term in the chiral 
expansion of the $D\bar D^*$ potential, as   formulated  above,
is justified for $\delta m_\pi\sim m_\pi\0$.   We assume that this framework might be still
applicable in the case of larger pion masses of the order of 300-400~MeV,
which is the typical value used in today's lattice QCD calculations in the charm
sector~\cite{lattice}.

It should be stressed that in the approach used here, besides the known $m_\pi$-dependences
of the masses and coupling constants, we also need to include  a dependence of the 
short-range interaction $C_0$ with an \emph{a priori} unknown strength,
in analogy to the investigations in the nucleon--nucleon (NN) sector
\cite{chiralextrapol_bonn,chiralextrapol_seattle}\footnote{See also a recent calculation of
\cite{Berengut:2013nh} where the pion mass dependence of NN contact interactions
was determined by resonance saturation using unitarised chiral
perturbation theory combined with lattice QCD results.}.
The inclusion of $m_\pi$-dependence of the short-range interaction is
necessary not only because of power counting arguments, but also demanded
by the renormalisation group: a $\Lambda$- and $m_\pi$-dependent
short-ranged operator is needed to absorb the $\Lambda$-dependence we encounter 
when varying the pion mass in the other quantities and especially in the pion propagator. 
Our formalism seems more complicated than that of \cite{Wang:2013kva}, where
it is claimed that a model-independent, parameter-free prediction can be given for 
the pion mass dependence of the $X$. However, it is our understanding that the result of \cite{Wang:2013kva} is
regularisation scheme as well as scale-dependent. 

We find that, as expected, the pion exchange itself gets weaker with increasing
pion mass. Meanwhile, the $m_\pi$-dependence of the counter term can now either enhance this feature, thus leading
to a rapid disappearance of the $X(3872)$ pole as $m_\pi$ is increased, or, on the contrary, weaken the
effect even that much, that the $X$ binding increases with increasing pion mass. In any case, the pion
mass dependence $E_B(m_\pi)$ turns out to be nontrivial. Thus our work provides important insights for the chiral
extrapolation in the $X(3872)$ sector. Furthermore, with the help of our findings and from the pion mass dependence 
of the $X(3872)$ pole  once it is available from lattice simulations, we will eventually be able to extract valuable
information on the physics of the interaction that leads to the formation of the $X(3872)$.

\section{Scattering equations}
\label{equations}

In this section we outline briefly our theoretical formulation
following closely the lines of \cite{Baru:2011rs}. 
The lowest-order $D^*D\pi$ interaction Lagrangian has the form \cite{Fleming:2007rp}
\begin{eqnarray}
{\cal L}=\frac{g_c}{\sqrt{2}f_\pi}\left({\bm D^*}^\dagger \cdot {\bm\nabla}\pi^a\tau^a D
+D^\dagger\tau^a{\bm\nabla}\pi^a\cdot {\bm D^*}\right),\quad
\pi=\left(
\begin{array}{cc}
\pi^0/\sqrt{2}&\pi^+\\
\pi^-&-\pi^0/\sqrt{2}
\end{array}
\right),
\label{lag}
\end{eqnarray}
where $g_c$ is the dimensionless $D^*D\pi$ coupling constant. The latter can be
fixed from the $D^{*0}\to D^0\pi^0$ width via the relation
\be
\varGamma(D^{*0}\to
D^0\pi^0)\equiv\varGamma_*=\frac{g_c^2m_0}{24\pi f_\pi^2m_{*0}}
\left[2\mu_q(D^0\pi^0)(m_{*0}-m_0-m_{\pi^0})\right]^{3/2},
\label{D*width}
\ee
where the reduced mass is defined as $\mu_q(XY)=m_Xm_Y/(m_X+m_Y)$. Here and in what follows, $m_0$, $m_c$,
$m_{*0}$, $m_{*c}$, $m_{\pi^0}$, and $m_{\pi^c}$ are the masses of the neutral
and charged $D$ mesons, $D^*$ mesons, and pions, respectively.

In the physical limit one has \cite{PDG} $f_\pi\0=92.4$ MeV,  
\begin{eqnarray*}
m_{\pi^0}\0=134.98~\mbox{MeV},\quad m_0\0=1864.84~\mbox{MeV},\quad m_{*0}\0=2006.97~\mbox{MeV},\\
m_{\pi^c}\0=139.57~\mbox{MeV},\quad m_c\0=1869.62~\mbox{MeV},\quad m_{*c}\0=2010.27~\mbox{MeV},
\end{eqnarray*} 
while the value $\varGamma_*\0=42$~keV can be deduced from the data for the charged $D^*$ decay modes \cite{PDG}. Then
relation (\ref{D*width}) gives $g_c\0=0.62$. 

Using the notation of \cite{Baru:2011rs}, the $P$-wave $D^*D\pi$ vertex can be written as
\be
v_{D^*D\pi}(\veq)=g\; {\bm \epsilon}\cdot \veq,
\label{g1}
\ee
where ${\bm \epsilon}$ is the $D^*$ polarisation vector and $\veq$ is the relative
momentum in the $D\pi$ system. The vertex coupling $g$ is related to the dimensionless constant 
${g_c}$ from the Lagrangian (\ref{lag}) as
\be\label{ggc}
g=\frac{g_c}{(4\pi)^{3/2}f_\pi}\left(\frac{m_0}{m_{*0}\mu_q(D^0\pi^0)}\right)^{1/2},
\ee
and its physical value is \cite{Baru:2011rs}
\be
g\0=1.29\cdot 10^{-5}~\mbox{MeV}^{-3/2}.
\ee
 
\begin{figure}[t]
\centerline{\epsfig{file=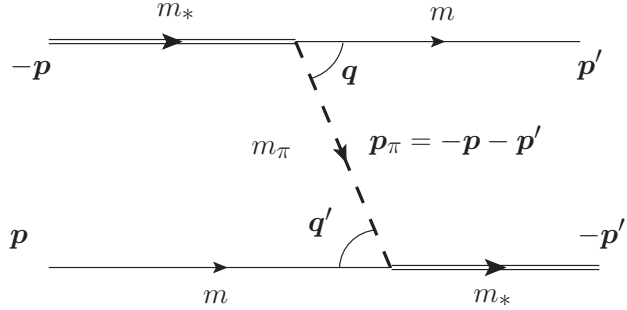,width=0.5\textwidth}}
\caption{Kinematics of the $D \bar D^*$ scattering due to the OPE. Double lines denote $D^*$'s,  single lines denote
$D$'s,  while the dashed line stands for  the pion.}\label{diagram}
\end{figure}

The OPE potential visualised in Fig.~\ref{diagram} is 
\be
V^{nn'}(\vep,\vep')=-g^2\frac{(\vep'+\alpha\vep)_n(\vep+\alpha\vep')_{n'}}{D_3(\vep,\vep')},
\quad \quad \alpha=\frac{m}{m+m_{\pi}},
\label{Vmn}
\ee
where the indices $n$, $n'$ are contracted with the corresponding indices of the $D^*$ polarisation vectors. 
The inverse three-body propagator reads
\be
D_3(\vep,\vep')=2m+m_\pi+\frac{p^2}{2m}+\frac{p'^2}{2m}+\frac{(\vep+\vep')^2}{2m_\pi}-M-i0.
\label{D3}
\ee
The OPE potential (\ref{Vmn}) interrelates the four $D$-meson channels defined as
\be
|0\rangle=D^0\bar{D}^{*0},\quad|\bar{0}\rangle=\bar{D}^0 D^{*0},\quad
|c\rangle=D^+ D^{*-},\quad|\bar{c}\rangle=D^- D^{*+}.
\ee
For the sake of convenience, we define the energy $E$ relative to the neutral two-body threshold:
\be
M=m_{*0}+m_0+E.
\label{ME}
\ee

Given the above definitions, the system of coupled Lippmann--Schwinger
equations for the $D\bar{D}^*$ $t$-matrix elements
$a_{00}^{nn'}(\vep,\vep')$ and $a_{c0}^{nn'}(\vep,\vep')$ in the
$C$-even channel has the form:
\bea
\left.
\begin{array}{l}
\hspace*{-2pt}a_{00}^{nn'}(\vep,\vep',E)=\lambda_0 V_{00}^{nn'}(\vep,\vep')
-\ds\sum_{i=0,c}\lambda_i\int\frac{d^3k}{\Delta_i(k)}V_{0i}^{nm}(\vep,\vek)a_{i0}^{mn'}(\vek,\vep',E),\\
\hspace*{-2pt}a_{c0}^{nn'}(\vep,\vep',E)=\lambda_c V_{c0}^{nn'}(\vep,\vep')
-\ds \ds\sum_{i=0,c}\lambda_i\int\frac{d^3k}{\Delta_i(k)}V_{ci}^{nm}(\vep,\vek)a_{i0}^{mn'}(\vek,\vep',E),
\end{array}
\right.
\label{aa}
\eea
where $\lambda_0=\langle 0|\vec\tau_1\cdot\vec\tau_2|\bar{0}\rangle=\langle
c|\vec\tau_1\cdot\vec\tau_2|\bar{c}\rangle=1$ and $\lambda_c=\langle
0|\vec\tau_1\cdot\vec\tau_2|\bar{c}\rangle=\langle
c|\vec\tau_1\cdot\vec\tau_2|\bar{0}\rangle=2$ are isospin factors 
for the $\pi^0$- and $\pi^\pm$-exchange, respectively. 
The inverse two-body propagators $\Delta_0$ and $\Delta_c$ take the form
\be
\Delta_0(p)=\frac{p^2}{2\mu_0}-E-\frac{i}{2}\varGamma_0(p),\quad
\Delta_c(p)=\frac{p^2}{2\mu_c}+\delta-E-\frac{i}{2}\varGamma_c(p),
\label{Deltas}
\ee
where $\delta=m_{*c}+m_c-m_{*0}-m_0$ and the ``running'' widths are
given by 
\bea
\varGamma_0(p)&=&\varGamma(D^{*0}\to D^0\gamma)+\frac{8\pi^2}{3}g^2
\sum_{k=0,c}\lambda_k\mu_{q kk}(q_{0kk}^3+i\tilde{\kappa}_{0kk}^3)\\
\varGamma_c(p)&=&\frac{8\pi^2}{3}g^2{
\sum_{j,k=0,c\;\&\;j\neq k}
}\lambda_k\mu_{q jk}(q_{cjk}^3+i\tilde{\kappa}_{cjk}^3). 
\eea
Here, 
$\mu_{qjk} \equiv\mu_q(D^j\pi^k)$ and the momentum
\be
q_{ijk}\equiv\sqrt{2\mu_{qjk}\left(M-m_i-m_j-m_{\pi^k}-\frac{p^2}{2\mu_{*i}}\right)},\quad i,j,k=0,c,
\ee
continues analytically below the corresponding threshold as $q_{ijk}\to i\kappa_{ijk}$. The constant shifts
\be
\tilde{\kappa}_{ijk}=\sqrt{2\mu_{qjk}(m_j+m_{\pi^k}-m_{*i})}\Theta(m_j+m_{\pi^k}-m_{*i})
\ee
cancel the contributions of the $\kappa_{ijk}$'s at thresholds and,
therefore, guarantee that the neutral and charged $D^*$ masses are defined as zeros of the real part of the inverse
propagators (\ref{Deltas}). This implies that, as the pion mass $m_\pi$ deviates from its physical value $m_\pi\0$, the
$D^*$ masses acquire contributions of the form:
\bea
\Delta m_{*0}&=&\frac12\mbox{Im}\left[\varGamma_0(p,m_\pi\0)-\varGamma_0(p,m_\pi)\right]|_{p=0,E=0},\nonumber\\[-3mm]
\label{dm0c}\\[-3mm]
\Delta m_{*c}&=&\frac12\mbox{Im}\left[\varGamma_c(p,m_\pi\0)-\varGamma_c(p,m_\pi)\right]|_{p=0,E=\delta}.\nonumber
\eea

The amplitudes $a_{ik}^{nn'}(\vep,\vep',E)$ can be represented as
\be
a_{ik}^{nn'}(\vep,\vep',E)=a_{ik}^{SS}(p,p',E)T_{SS}^{nn'}+a_{ik}^{DS}(p,p',E)T_{DS}^{nn'}
\ee
with the help of the $S$- and $D$-wave projectors 
\be
T_{SS}^{nn'}=\frac{1}{4\pi}\delta_{nn'},\quad T_{DS}^{nn'}=\frac{1}{4\pi\sqrt{2}}
(\delta_{nn'}-3n_nn_{n'}).
\ee
For the quantum numbers $1^{++}$ and for the neutral mesons in the final state, we are interested only in the
$a_{00}^{SS}$. A similar decomposition is done for the potential $V_{ik}^{nn'}$. 
Then, in order to describe the short-range dynamics, we modify
the component of the potential $V_{ik}^{SS}$ as
\be
V_{ik}^{SS}(p,p')\to C_0+V_{ik}^{SS}(p,p'),
\ee
where the constant contact interaction $C_0$ is tuned to produce a bound state at $E=-E_B=-0.5$~MeV. It has
to be noticed that,
because of the $P$-wave vertices, the integrals in the Lippmann--Schwinger equation (\ref{aa}) diverge.
We regularise them by a sharp cut-off $\Lambda$ and assume a proper $\Lambda$-dependence of the contact
interaction $C_0(\Lambda)$ which ensures that physical observables are
(approximately) $\Lambda$-independent. This program was successfully
carried out in \cite{Baru:2011rs} where all further details can be
found. We only stress here that our approach explicitly preserves unitarity and treats OPE nonperturbatively. 

\section{Masses, decay constants, and widths}\label{masseswidths}

For the sake of convenience, we introduce the ratio
$m_\pi/m_\pi\0\equiv\xi$ and describe the ``running'' behaviour of all physical quantities
entering the problem in terms of this ratio $\xi$. Physical values are
readily restored for $\xi=1$. 

The $m_\pi$-dependence of the $D$- and $D^*$-meson masses was studied
in a recent paper \cite{Cleven:2010aw} within the framework of unitarised $SU(3)$ chiral
perturbation theory. The tree-level expressions for $m(\xi)$ and
$m_*(\xi)$ corresponding to the expansion about the $SU(2)$ chiral
limit with the charm quark mass fixed at its physical value have the form 
\be
m(\xi)=m\0\left[1+h_1\left(\frac{m_\pi\0}{m\0}\right)^2(\xi^2-1)\right],\quad
m_*(\xi)=m_*\0\left[1+h_1\left(\frac{m_\pi\0}{m_*\0}\right)^2(\xi^2-1)\right]+\Delta m_*,
\label{mDs}
\ee
where $h_1\approx 0.42$, the shift $\Delta m_*$ is given in
(\ref{dm0c}) and, in order to simplify the presentation, we do not
distinguish between neutral and charged states. 

For the pion decay constant, we resort to the one-loop chiral
perturbation theory result (see, for example, \cite{Gasser:1983yg}) which in our notation has the form
\be
f_\pi(\xi)=f_\pi\0\left[1+\left(1-\frac{f}{f_\pi\0}\right)(\xi^2-1)-\frac{\oo{m_\pi}}{8\pi^2f^2}\xi^2\ln\xi\right]
\label{fpi2}.
\ee
Notice that we have expressed the corresponding low-energy constant
$l_4$ in terms of the pion decay constant $f$ in the chiral limit for
which we adopt the value~\cite{Becirevic:2012pf}
\be
f_\pi(m_\pi=0)\equiv f=85~\mbox{MeV}.
\label{f85}
\ee

To establish the $m_\pi$-dependence of the $D^*D\pi$ coupling $g$ we use the lattice results of
\cite{Becirevic:2012pf} (in particular, the fit ChPT-II which incorporates one-loop effects). In order to avoid
confusion with notations, we use the subscript ``${\rm lat}$'' for some parameters from \cite{Becirevic:2012pf}. We
also re-define the pion decay constant of \cite{Becirevic:2012pf} to comply with the standard convention
(see, in particular, (\ref{f85})). As follows from (\ref{ggc}),
\be\label{gmpi}
g=g\0\left(\frac{m_\pi\0}{m_\pi}\right)^{1/2}
\left(\frac{m_*\0}{m_*}\right)^{1/2}
\left(\frac{m+m_\pi}{m\0+m_\pi\0}\right)^{1/2}
\left(\frac{f_\pi\0}{f_\pi}\right)
\left(\frac{g_c}{g_c\0}\right),
\ee
with the ratio $g_c/g_c\0$ as a function of $\xi$ extracted from \cite{Becirevic:2012pf} in the form:
\bea
\frac{g_c}{g_c\0}&=&1+C_1(\xi^2-1)+C_2\; \xi^2\ln\xi,\\
C_1&=&1-\left(1-\frac{(1+2g_0^2)}{8\pi^2}\left(\frac{m_\pi\0}{f}\right)^2\ln\frac{m_\pi\0}{\mu_{\rm lat}}+\alpha_{\rm
lat} \oo{m_\pi}\right)^{-1},\\
%C_2&=&\left.\left(C_1-(1-C_1)\alpha_{\rm lat}\oo{m_\pi}\right)\right/\ln\frac{m_\pi\0}{\mu_{\rm lat}}.
%,\\
C_2&=&  -\frac{(1+2g_0^2)}{8\pi^2}\left(\frac{m_\pi\0}{f}\right)^2  
\left(1-\frac{(1+2g_0^2)}{8\pi^2}\left(\frac{m_\pi\0}{f}\right)^2\ln\frac{m_\pi\0}{\mu_{\rm lat}}+\alpha_{\rm
lat} \oo{m_\pi}\right)^{-1}.
\eea

Further, the values of the various parameters taken from \cite{Becirevic:2012pf} are:
\be
g_0=0.46,\quad\alpha_{\rm lat}=-0.16~\mbox{GeV}^{-2},\quad \mu_{\rm lat}=1~\mbox{GeV}.
\ee

\begin{figure}[t]
\centerline{
\epsfig{file=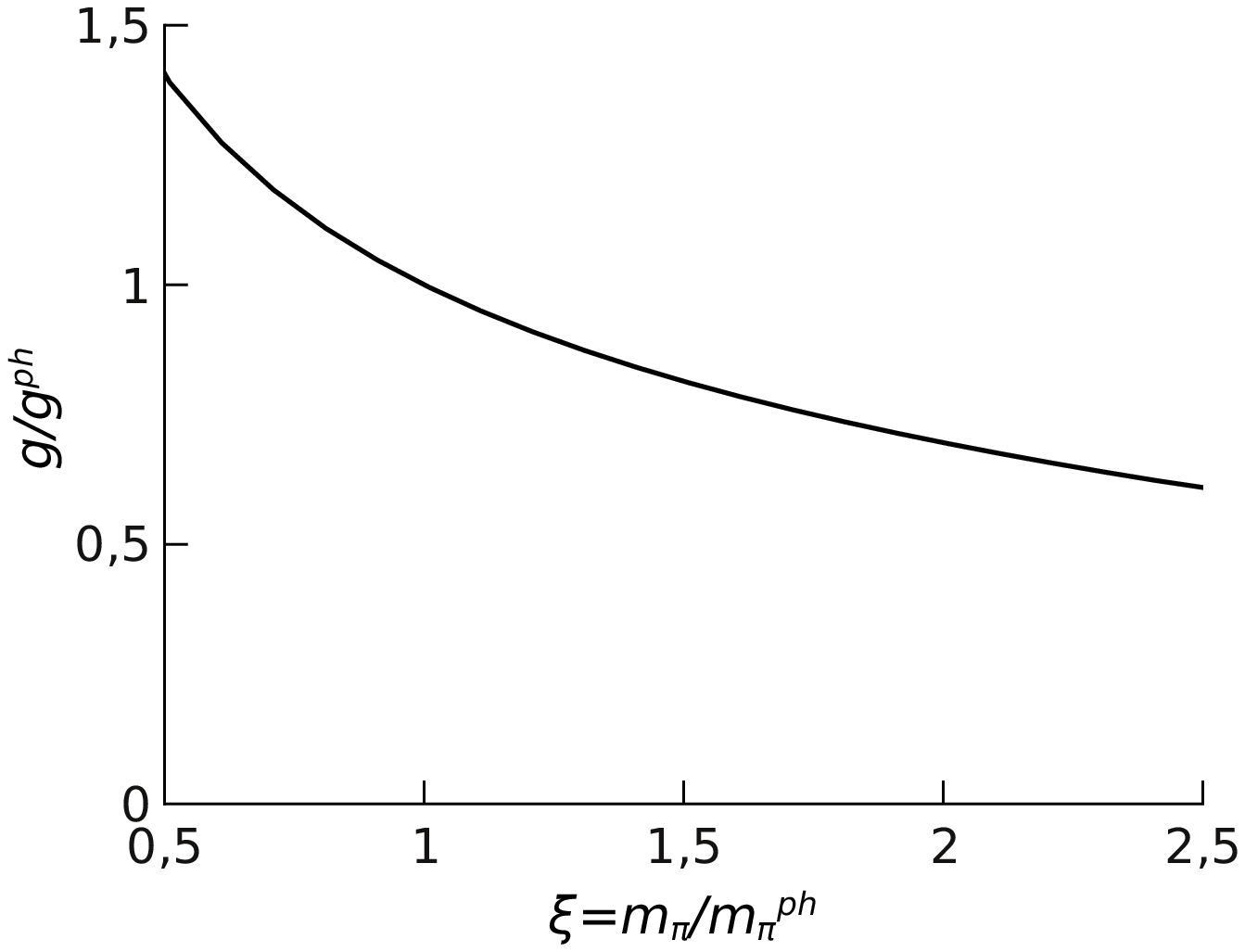,width=0.35\textwidth}\hspace*{-5mm}
\epsfig{file=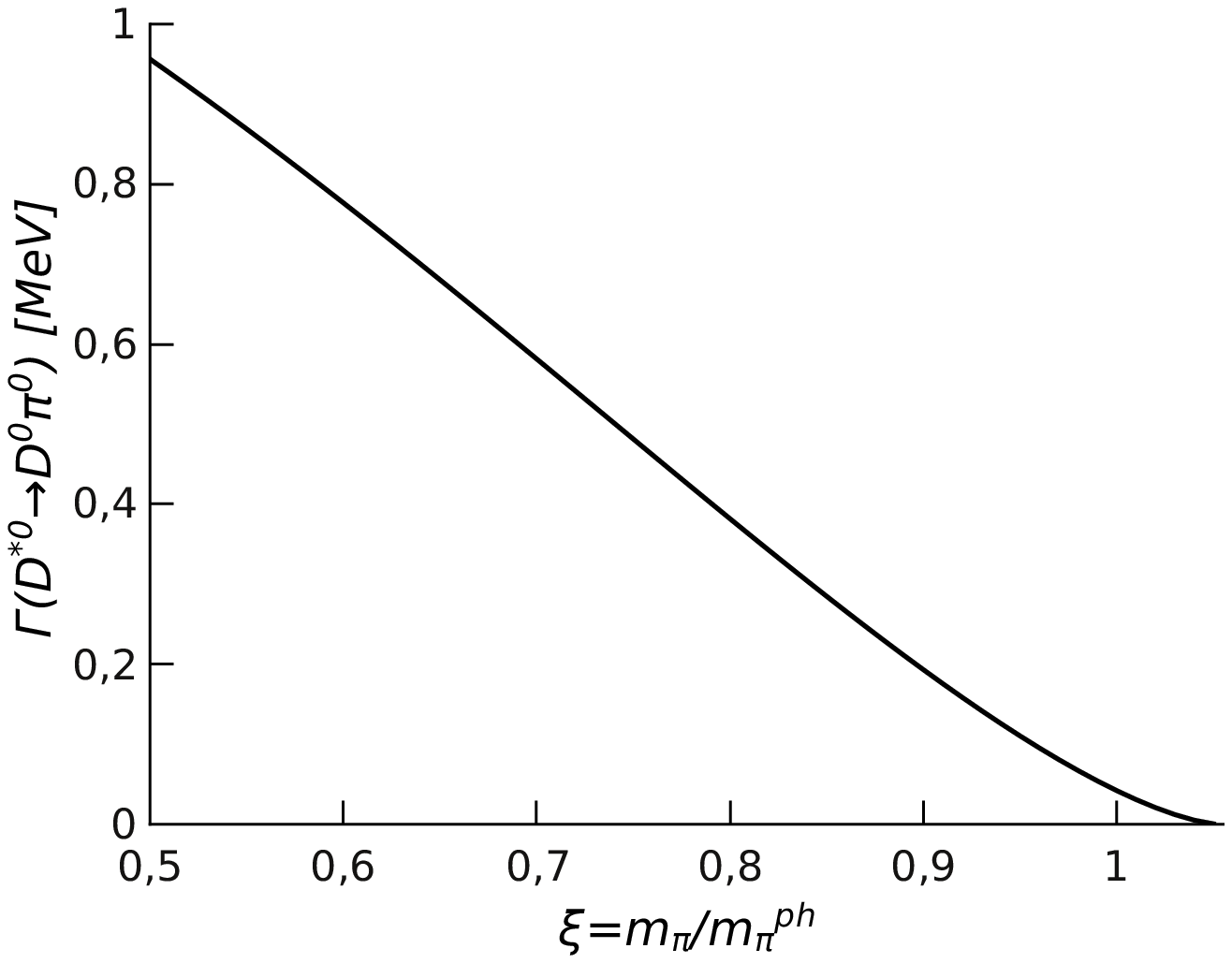,width=0.35\textwidth}\hspace*{-6mm}
\epsfig{file=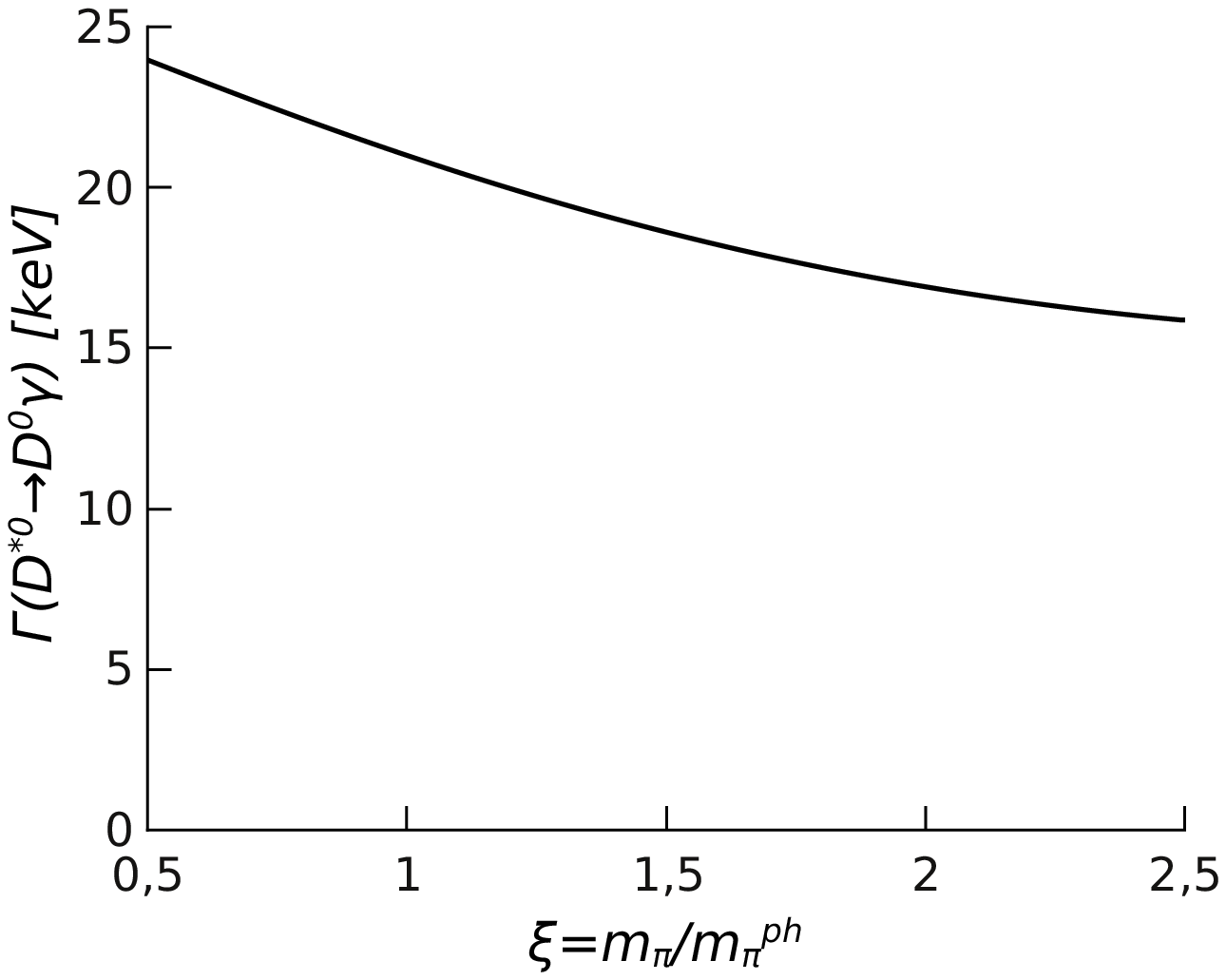,width=0.35\textwidth}
}
\caption{Dependence of the ratio $g/g\0$ (left panel), the
$D^{*0}\to D^0\pi^0$ width (middle panel) and the $D^{*0}\to
D^0\gamma$ width (right panel) on the varying pion mass $\xi=m_\pi/m_\pi\0$.}
\label{figwidth}
\end{figure}

In Fig.~\ref{figwidth} we give the $m_\pi$-dependence of the coupling
constant $g$ (left panel) and that of the $D^{*0}\to
D^0\pi^0$ decay width (middle panel). The width
$\varGamma(D^{*0}\to D^0\pi^0)$ vanishes already at $m_\pi/m_\pi\0\approx
1.05$ 
because of the closing phase space. The coupling
constant $g$ does not vanish at this point, however, and defines the
analytical continuation of the loop operator below threshold. 

For the $D^*\to D\gamma$ width, we adopt the results derived
in \cite{Amundson:1992yp}, namely 
\be
\varGamma(D^{*0}\to D^0\gamma)\equiv\varGamma_\gamma=\frac13\alpha_{\rm EM}|\mu_1|^2k_\gamma^3,
\label{Gg}
\ee
where $\alpha_{\rm EM}=1/137$ is the fine structure constant, $k_\gamma$ is the
photon momentum, and $\mu_1$ is the transition magnetic
moment which depends on the pion mass. Notice that in
\cite{Amundson:1992yp}, in addition to pion loops, also effects of the kaon loops were considered in order to determine
the quark mass dependence of $\mu_1$. In the framework of SU(2) chiral
perturbation theory, any $m_\pi$-dependence emerging through
the kaon mass is represented by local,
$m_\pi^2$-dependent operators in the effective
Lagrangian. At the order in the $\xi$-expansion we are working, the
contributions from the kaon loops to the $m_\pi$-dependence of 
$\mu_1$ can be neglected. The resulting dependence of
$\varGamma_\gamma$ on $m_\pi$ is visualised in the right panel of
Fig.~\ref{figwidth}, 
while the physical value deduced from the charged $D^*$ decays modes is $\varGamma_\gamma\0=21~\mbox{keV}$.
Taking into account the effects of kaon loops from \cite{Amundson:1992yp} leads to a sort of resonance
saturation of the corresponding higher-order terms that gives rise to an extra $ m_\pi$-dependence of
$\varGamma_\gamma$.
We have verified  that the resulting dependence $\varGamma_\gamma(\xi)$ is nearly
indistinguishable from the one shown in Fig.~\ref{figwidth}.

\section{The contact interaction}\label{ctsec}

In this section we discuss in detail the derivation of the dependence of the contact interaction strength, $C_0$, 
on the pion mass. As was
explained above, this term incorporates  the unknown short-range  dynamics  
present in the system. The nature of the latter part is difficult to determine, so we are 
forced to resort to an educated guess on the
relevant dependence $C_0(m_\pi)$ based on (i) the assumption that the $X$ is a bound state and (ii) the requirement
that the binding energy $E_B(m_\pi)$ is $\Lambda$-independent (that is, one is able to renormalise the full theory with
dynamical pions at least in the leading order in $\Lambda$). Thus we postulate for the ``running'' contact interaction
the form
\be
C_0(\Lambda,\xi)=C_0\0+\delta C_0=C_0\0(\Lambda)\left(1+f(\Lambda)\frac{\delta m_\pi^2}{\rm
M^2}\right)=C_0\0(\Lambda)\left(1+f(\Lambda)\frac{\oo{m_\pi}}{\rm M^2}(\xi^2-1)\right),
\label{C0run}
\ee
where, as it was explained in the introduction, the expansion is done
in terms of the ratio of the small scale $\delta m_\pi$ to the large
scale ${\rm M}$, and only the leading term is kept. The leading $\Lambda$-dependence of the contact interaction is
captured by the physical-limit quantity $C_0\0(\Lambda)$, while the dimensionless function
$f(\Lambda)$ absorbs the extra $\Lambda$-dependence which appears for values
of the pion mass away from the physical point. Therefore, we fix the $\Lambda$-dependence of the contact interaction
requiring that both the binding energy $E_B$ itself as well as its slope at the physical point, 
$(\partial E_B/\partial m_\pi){\big |}_{m_\pi=m_\pi\0}$, are $\Lambda$-independent. 

Here and in what follows  we adopt the value of M $\sim m_{\rho}\sim 800$ MeV for the relevant hard scale in the problem
with dynamical pions. Further, since $\Lambda$ has to be larger than the maximal pion mass considered in the study, $
m_\pi \simeq$ 300 MeV, but smaller than the hard scale M, we let the cut-off $\Lambda$ vary in the range of $400 \ldots
700$~MeV.

In order to get a better insight for the $C_0(\Lambda,\xi)$ and, in particular,
investigate properties of the function $f(\Lambda)$, we first consider an analytically solvable model with the
interaction given by the
contact potential $C_0$ only \cite{Baru:2011rs}. We also stick to the isospin limit for simplicity. We start from the
physical pion mass.
The system of Lippmann--Schwinger equations (\ref{aa}) reduces to two decoupled equations for the amplitudes
$a_0=(a_{00}-a_{c0})/2$ and $a_1=(a_{00}+a_{c0})/2$ which correspond to the isosinglet and isotriplet channels,
respectively. In the isoscalar channel relevant for the $X$,
the equation reads:
\be
a_0=-C_0+C_0 I(E) a_0,
\label{a0a1}
\ee
where
\be
I(E)=\int_0^{\Lambda} ds \frac{s^2}{s^2/(2\mu)-E}\approx 2\mu\Lambda-
\pi\mu\sqrt{-2\mu E}+O\left(\frac{\sqrt{-2\mu E}}{\Lambda}\right),\quad \mu=\frac{mm_*}{m+m_*}.
\label{IE}
\ee
The system possesses a bound state in the isosinglet channel at $E=-E_B$ if 
\be
C_0I(-E_B)=1,
\label{beq}
\ee
as  follows from  (\ref{a0a1}). This means that 
while both $C_0$ and $I(-E_B)$ are strongly $\Lambda$-dependent, their
product is a renormalisation group invariant $\Lambda$-independent quantity, and the bound-state equation (\ref{beq})
defines the contact interaction to be $C_0(\Lambda)=I^{-1}(-E_B,\Lambda)$. 

If the pion mass deviates from $m_\pi\0$, the corresponding expansion of $C_0$ near its physical value reads:
\be
C_0=I^{-1}(-E_B)\approx I^{-1}(-E_B\0)
+\frac{\partial I^{-1}(-E_B\0)}{\partial\mu\0}\frac{\partial\mu\0}{\partial m_\pi^2}\delta m_\pi^2
+\frac{\partial I^{-1}(-E_B\0)}{\partial E_B\0}\frac{\partial E_B\0}{\partial m_\pi^2}\delta m_\pi^2
\label{CIrun}
\ee
$$
=C_0\0\left[1-\left(\frac{\partial \ln I(-E_B\0)}{\partial\mu\0}\frac{\partial\mu\0}{\partial m_\pi^2}
+\frac{\partial \ln I(-E_B\0)}{\partial E_B\0}\frac{\partial E_B\0}{\partial m_\pi^2}\right)\delta m_\pi^2\right].
$$
It can be re-written in the form of (\ref{C0run}) with
\be
\frac{f(\Lambda)}{{\rm M}^2}=-\frac{h_1}{(m^{\rm{ph}})^2}+
\frac{\pi\mu\0}{2\Lambda\sqrt{2\mu\0 E_B\0}}\frac{\partial E_B\0}{\partial m_\pi^2},
\label{xiC}
\ee
where we used that
\be
\frac{\partial\ln I(-E_B)}{\partial E_B}\approx-\frac{\pi\mu}{2\Lambda\sqrt{2\mu E_B}},\quad
\frac{\partial\ln I(-E_B)}{\partial\mu}\frac{\partial\mu}{\partial m_\pi^2}=h_1\left(
\frac{1}{m^2}-\frac{1}{mm_*}+\frac{1}{m_*^2}\right)\approx\frac{h_1}{m^2},
\label{C0mu}
\ee
and the difference between the $m_*$ and $m$ is neglected. The
constant $h_1$ is given at the beginning of Sect.~\ref{masseswidths}. 
Equation (\ref{xiC}) yields an analytical expression for the variation of the coefficient $f$ with the cut-off $\Lambda$
which keeps the slope $(\partial E_B/\partial m_\pi)|_{m_\pi=m_\pi\0}$ in the contact theory independent of $\Lambda$. 
In the theory with dynamical pions, the same requirement that the binding energy slope at the physical point is
$\Lambda$-independent allows one to determine the cut-off dependence $f(\Lambda)$ numerically.

\begin{figure}[t]
\centerline{\epsfig{file=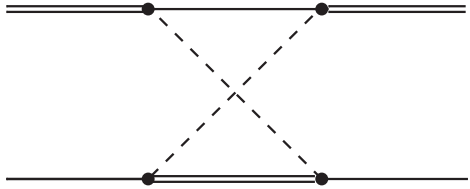, width=0.4\textwidth}}
\caption{Two-pion exchange contribution. Double (single) lines denote $D^*$'s ($D$'s) while dashed lines denote
pions. \label{TPEdia}}
\end{figure}

Finally, in order to implement (\ref{C0run}) in the calculation of the binding energy 
$E_B$ as a function of the pion mass $ m_\pi$, we need an estimate for $f(\Lambda)$.
To this end we consider an example of a physical mechanism which is not included explicitly in our calculations and
estimate its contribution to the $m_\pi$-dependence of $C_0$. Specifically, we consider the crossed-box
two-pion-exchange (TPE) diagram shown in Fig.~\ref{TPEdia}. The corresponding amplitude is $M_{\rm TPE}\sim ({
g_c}\sqrt{mm_*}/f_\pi)^4\times I_{\rm TPE}$, with
\be 
I_{\rm TPE}=\int\frac{d^4 q}{(2\pi)^4}\frac{\veq^4}{(q^2- m_\pi^2+i0)^2((p-q)^2-m_*^2+i0)((p_*-q)^2-m^2+i0)},
\label{TPE}
\ee
where $p=(m,{\bm 0})$ and $p_*=(m_*,{\bm 0})$ are the 4-momenta of the initial (and final) $D$ and $D^*$
meson, respectively. We now estimate the contribution of the TPE diagram to $f(\Lambda)$ with the help of a naive
dimensional analysis (NDA). This can be most easily achieved in the framework of time-ordered perturbation theory:
\be
I_{\rm TPE}\approx -i \int\frac{d^3 q}{(2\pi)^3} \frac{1}{(2\omega)(2\omega_*)(2\omega_{\pi})^2}  
\frac{\veq^4}{(m+m_*-2\omega_{\pi}-\omega-\omega_*) 
(m_*-\omega-\omega_{\pi}+i0)^2},
\ee
where $\omega_i=\sqrt{\veq^2+m_i^2}$ are the energies of the $\pi$-, $D$-, and $D^*$-meson and we keep only the dominant
time ordering with two $\pi DD$ and one $\pi\pi DD^*$ intermediate states. 
There are, in general, three scales contributing to the integral: (i) $q\sim\Lambda$, (ii) $q\sim m_\pi$, and (iii)
$q\sim\sqrt{2 m_\pi(m_*-m- m_\pi)}$ (due to the $\pi DD$ cuts). The relevant short-range contribution $\propto m_\pi^2$
stems only from the first scale  after expanding the integral to the order $ m_\pi^2/\Lambda^2$  and from the second
scale, both yielding $ m_\pi^2/{[(4\pi)^2 m m_*]}$ for the relevant part of the integral. 

Thus, relative to $C_0\0$, the $m_\pi$-dependent part of the short-range interaction $\delta C_0$ induced by the TPE
contribution can be estimated as
\be
\frac{\delta C_0}{C_0\0} \sim \frac{ M_{\rm TPE}}{M_{\rm OPE}} \sim 
 g_c^2 \frac{\delta m_\pi^2}{(4\pi f_\pi)^2 }  \sim 
\left(  g_c^2  \frac{\rm M^2}{{(4\pi f_\pi)^2 } }\right) \frac{\delta m_\pi^2}{\rm M^2}, 
\label{dCest}
\ee
where we used the fact that the contact term $C_0\0$ is saturated by the OPE amplitude at large momenta, $M_{\rm
OPE}\sim ({ g_c}\sqrt{mm_*}/f_\pi)^2$. Matching (\ref{dCest}) and (\ref{C0run}), with M~$\sim 4\pi f_\pi$ and
${ g_c}\sim 1$, gives $f(\Lambda) \sim 1$. 

\begin{figure}[t]
\centerline{\epsfig{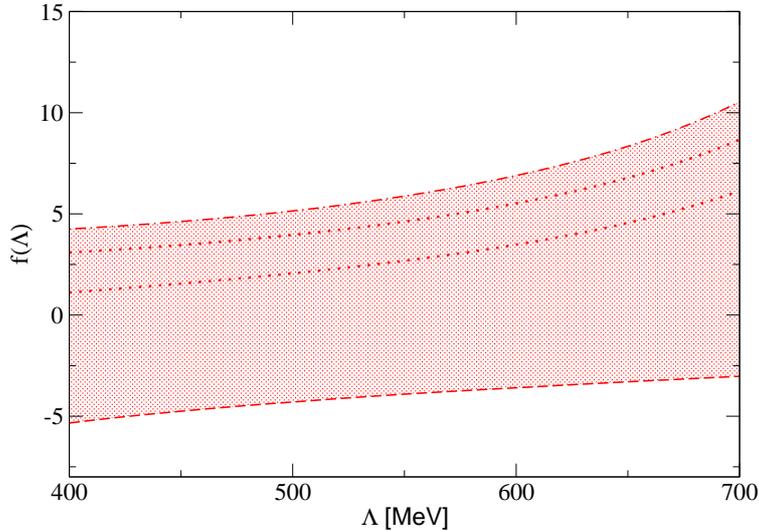}}
\caption{Cut-off dependence of the quantity $f(\Lambda)$ for various values of the slope 
$(\partial E_B/\partial m_\pi){\big |}_{m_\pi=m_\pi\0}$. The red filled band indicates the values of $f(\Lambda)$
regarded as natural. The narrow band bounded by the dotted lines corresponds to the natural range for $f(\Lambda)$
derived from the contact theory.}\label{flam}
\end{figure}

The estimate for $f(\Lambda)$ sets a constraint on the magnitude of the slope of the binding energy (see, for example,
(\ref{xiC}) for the case of the contact theory). In particular, for $f(\Lambda)\sim 1$ the shift of the binding
energy $\delta E_B\sim E_B$ for $\delta m_\pi\sim m_\pi$ can be interpreted as natural. In Fig.~\ref{flam}, as dotted
curves, we show the trajectories $f(\Lambda)$ which correspond to the (cut-off-independent) slope of 
$(\partial E_B/\partial m_\pi){\big |}_{m_\pi=m_\pi\0}=\pm E_B\0/ m_\pi\0=\pm 0.4\times 10^{-2}$. One can see that, at
these trajectories $f(\Lambda)\sim 1$, so that, indeed, the estimates $f(\Lambda)\sim 1$ and $\delta E_B/\delta
m_\pi\sim E_B/m_\pi$ are reasonably consistent with each other.

Given that other mechanisms, not considered here, can potentially affect $f(\Lambda)$ too, we use a more conservative
estimate for $f(\Lambda)$ and regard as natural its values lying in the range
$-5\ldots +5$. This leads to the broad filled band shown in Fig.~\ref{flam}. Then, as the pion mass varies in the range
$ m_\pi\0\leqslant m_\pi\leqslant 2 m_\pi\0$, the change in the contact term $\delta C_0$ in (\ref{C0run}) does not
exceed
50\% of $C_0\0$.

To summarise, in the full problem which incorporates dynamical pions,
we treat as natural the dependence of the contact interaction
on $m_\pi$ as given by (\ref{C0run}) and adopt a
conservative estimate for $f(\Lambda)$, as shown by the filled red band in Fig.~\ref{flam}.
The explicit form of $f(\Lambda)$ is determined by demanding the slope 
$(\partial E_B/\partial m_\pi){\big |}_{m_\pi=m_\pi\0}$ to be independent of $\Lambda$, the latter requirement to be
understood as a numerical implementation of the renormalisation group.

\section{Results and discussion}

\begin{figure}[t]
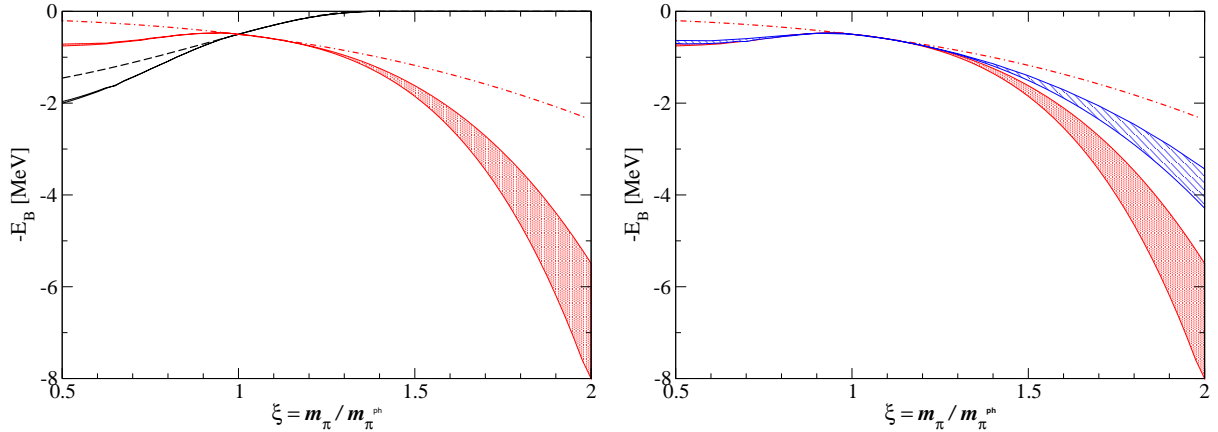

\centerline{
\epsfig{file=FIGS.DIR/EB_Full_pionless_paper.eps, width=0.47\textwidth}\hspace*{2mm}
\epsfig{file=FIGS.DIR/EB_Full_pionless_saturation.eps, width=0.47\textwidth}
}
\caption{The pion mass dependence of the binding energy 
for various values of the slope $(\partial E_B/\partial m_\pi){\big |}_{m_\pi=m_\pi\0}$ and for different cut-offs. 
Left panel: the red filled band corresponds to 
$(\partial E_B/\partial m_\pi){\big |}_{m_\pi=m_\pi\0}= 0.7\times 10^{-2}$ (the upper bound of the
allowed $f(\Lambda)$ shown by the dot-dashed line in Fig.~\ref{flam}) for the cut-off variation in the range 
$\Lambda\in [400,700]$ MeV; the black filled band (that looks almost like a single curve) is the same
but for the negative slope $(\partial E_B/\partial m_\pi){\big |}_{m_\pi=m_\pi\0}=-1.5\times 10^{-2}$. Right panel: the
pion mass dependence of the binding energy for the slope $(\partial E_B/\partial m_\pi){\big
|}_{m_\pi=m_\pi\0}=0.7\times 10^{-2}$ and for 
different cut-offs, namely, red filled band for $\Lambda\in [400,700]$ MeV, as in the left panel; blue
hatched band for 
unnaturally low cut-offs $\Lambda\in [150,250]$ MeV. The dashed and dash-dotted curves in both panels correspond to the
results of the pure contact theory.
}\label{results}
\end{figure}

As explained in the previous section, the functional form of the $m_\pi$-dependent short-range term $f(\Lambda)$ is
determined by the requirement that the slope $(\partial E_B/\partial m_\pi){\big |}_{m_\pi=m_\pi\0}$ is
cut-off-independent.
Based on our conservative estimate, we allow for those values of $f(\Lambda)$ that, for
$400~\mbox{MeV}\leqslant\Lambda\leqslant 700~\mbox{MeV}$, lie within the red filled band in Fig.~\ref{flam}. The
corresponding allowed values for the slope are $(\partial E_B/\partial m_\pi){\big |}_{m_\pi=m_\pi\0}=(-2.2 \ldots
0.7)\times 10^{-2}$, where the lowest value corresponds to the dashed curve in Fig.~\ref{flam} (lower bound)
whereas the largest one corresponds to the dot-dashed curve (upper bound). Results for the pion mass dependence of the
$X(3872)$ binding energy are shown in Fig.~\ref{results}.
As the pion mass increases, the pion exchange gets weaker. Therefore, in the absence of 
additional short-range forces ($f(\Lambda)\equiv 0$), the binding energy would smoothly decrease with the increase of
the pion mass. 

For negative values of $f(\Lambda)$, the contact interaction interferes constructively with the OPE contribution
resulting in a more rapid fall of $E_B$. This scenario is exemplified in the left panel of Fig.~\ref{results} by the
black band
corresponding to $(\partial E_B/\partial m_\pi){\big |}_{m_\pi=m_\pi\0}=-1.5\times 10^{-2}$ and $\Lambda\in
[400,700]$~MeV. 
The larger is $|f(\Lambda)|$, the quicker the $X(3872)$ pole moves towards the threshold, where it then turns into a
virtual state that is observable on the lattice only via an unnaturally large attractive scattering length, as long as
it is located very close to the threshold. If this scenario were realised in nature, it would mean that the
chiral dynamics of the $X$ is  dominated by short-range mechanisms as follows from matching with the 
results of the pure contact theory depicted by the dashed black curve: only in the region of small pion masses
$\xi\lesssim 0.9$ do the results of the pionful and pionless calculations start to deviate from
each other. 

On the contrary, for relatively large and positive values of $f(\Lambda)$, the short-range contributions
from the OPE and $f(\Lambda)$ interfere destructively so that pion dynamics starts playing an important role.
Specifically, for $(\partial E_B/\partial m_\pi){\big |}_{m_\pi=m_\pi\0}= 0.7\times 10^{-2}$ (the upper bound of 
the allowed region of $f(\Lambda)$ in Fig.~\ref{flam}), one observes a
sizeable deviation of the predictions of the theory with dynamical pions (red filled band in 
the left panel of Fig.~\ref{results}) from the results based on the pure contact interactions (red dot-dashed curve).

Further, in the right panel of Fig.~\ref{results} we show how the results of the full dynamical theory saturate those
based on the pure contact interactions. To this end we use unnaturally low cut-offs $\Lambda\in [150,250]$ MeV,
that result in integrating out pionic degrees of freedom to a large extent, and observe, as expected, that the
predictions of the full theory with dynamical pions converge to those for the pure contact theory,  as  formulated in
Sec.\ref{ctsec} (compare the blue  hatched
and the red filled bands in Fig.~\ref{results} versus the red dash-dotted curve). 
On the other hand, the deviation of the blue band from the dash-dotted curve can be accounted for by the
fact that the pion mass dependence stemming from the OPE, as provided by the propagator
$D_3( m_\pi)$ (see (\ref{D3})) and the coupling constant $g( m_\pi)$ (see (\ref{gmpi})) is more rich than just the
$ m_\pi^2$-variation of the contact term, as given by (\ref{C0run}).

Finally, it should be understood that the $m_\pi$-dependence of the
$X(3872)$ mass also receives contributions from $m_\pi$-dependence of
the masses of the $D$ and $D^*$ mesons. While these contributions
can be straightforwardly calculated in the framework of chiral
perturbation theory and taken into account in the analysis of $M_X( m_\pi)$,
they are irrelevant for the analysis of 
the binding energy of the $X(3872)$, that is, the energy
relative to the neutral two-body threshold $m_{*0} + m_0$. The $m_\pi$-dependence of the binding energy of
the $X(3872)$ induced by the $m_\pi$-dependence of $m_{*}$ and $m$ in
the kinetic energy terms entering the two- and three-body propagators
is numerically small and lies beyond the accuracy of our analysis. 

To summarise, the pion mass dependence of the binding energy of the $X(3872)$
analysed in our work appears to be highly nontrivial. Future lattice QCD calculations are expected to be
able to distinguish between different scenarios considered in our work and might provide
further insights into the chiral dynamics of the $X$.

\section{Summary}

In this work the pion mass dependence of the $X(3872)$ was studied at leading
order in chiral effective field theory. To achieve cut-off independence both
at the physical pion mass as well as for unphysical pion masses, it was
necessary to take into account a cut-off dependence of both the $m_\pi$-independent
and $m_\pi$-dependent parts of a $D\bar{D}^*$ contact interaction.
Then while the former is determined by fixing the binding energy of the $X$ at the physical point
when varying the regulator, the latter involves, at the order we are working, 
one unknown constant. This constant was expressed in terms of the slope of the binding energy of the $X$ at the
physical point which 
is, in principle, measurable in lattice QCD.
Depending on the employed value of the slope, different scenarios for the  
binding energy of the $X(3872)$ were found to be possible ranging from its
quick disappearance to an increasingly deeply bound state when the pion mass is increased. Thus, once the pion mass
dependence of the binding energy of the $X(3872)$ is determined in lattice QCD,
the $m_\pi$-dependence of the $D\bar{D}^*$
scattering potential can be deduced from our work leading to nontrivial insights into the binding mechanism of the
$X(3872)$. 
 
At the same time, our work demonstrates that the pion mass dependence of the $X(3872)$ binding energy is nontrivial for
all scenarios and 
provides a solid basis for chiral extrapolations of this important quantity. 
We further emphasise that in order to determine the $ m_\pi$-dependence of the $X(3872)$ mass $M_X=m_0+m_{*0}-E_B$, one
needs to know accurately the pion mass dependence of the neutral $D\bar{D}^*$ threshold. 
While it can be straightforwardly calculated in the framework of chiral perturbation theory,
we believe that the most accurate determination of the binding mechanism of the $X(3872)$ on the lattice would 
require a simultaneous measurement of the $X(3872)$ mass and the $D\bar{D}^*$ threshold.
\bigskip

The authors are grateful to to F.-K. Guo for valuable comments.
This work is supported in part by the DFG and the NSFC through funds provided
to the Sino-German CRC 110 ``Symmetries
and the Emergence of Structure in QCD'', by the
EU Integrated Infrastructure Initiative HadronPhysics3 and the ERC project 259218 NUCLEAREFT.

\end{document}